\documentclass[twocolumn,pra]{revtex4}
\usepackage{graphicx}
\usepackage{dcolumn}
\usepackage{bm}
\begin{document}

\preprint{APS/123-QED}

\title{Decoherence of collective atomic spin states due to inhomogeneous coupling}
\author{C. P. Sun$^{1,2}$, S. Yi$^1$, and L. You$^1$}
\affiliation{$^1$School of Physics, Georgia Institute of
Technology, Atlanta, GA 30332, USA}

\affiliation{$^2$Institute of Theoretical Physics, The Chinese
Academy of Sciences, Beijing 100080, People's Republic of China}
\date{\today}

\begin{abstract}
We investigate the decoherence of a superposition of
symmetric collective internal states of an atomic ensemble due to
inhomogeneous coupling to external control fields.
For asymptotically large system, we find the characteristic
decoherence rate scales as $\sqrt{N}$ with $N$ being the total
number of atoms. Our results shed new light
on attempts for quantum information processing
and storage with atomic ensembles.
\end{abstract}

\pacs{03.65.Yz, 03.67.Lx, 03.75.Fi}

\maketitle

Coherent quantum information encoding and processing has recently emerged as
a major goal for the physics community. Despite the seemingly insurmountable
difficulties, a rich variety of implementations are being
pursued in laboratories across the globe. Among the early success is the
apparent ability in simulating quantum operations with liquid based
NMR \cite{nmr1,nmr2}. Recent theoretical
efforts indicate, however, that in the pseudo-pure state
approach using NMR, {\it quantum entanglement}, a key
element for powerful quantum information processing, was in fact no present
\cite{caves,maciek}. Room temperature NMR technique, is therefore limited
and can not be explored fully to benefit from an exponentially large Hilbert
space of only a polynominally scaled resources
and controls \cite{nmr}. Nevertheless, early NMR based
experiments have provided useful insight into the operations of genuine
quantum computers \cite{rf}.

Over the last few years, using symmetric collective internal states
of an atomic ensemble, has attracted much attention
\cite{lukin0,lukin1,Flei0,Flei1,zol0}.
The pros and cons of such an approach assisted with cavity photon-atom
interaction was recently discussed by Fleischhauer {\it et al.}
\cite{Flei0,Flei1}.
In normal cavity QED based quantum
computing implementations, atomic qubits are entangled and logic operations
performed through their interaction with the common cavity photon
quantum field. To maintain quantum coherence, it is important to reach
the so-called strong coupling regime, when the single-photon coherent
coupling $g_0\gg \gamma,\kappa $, the atomic and cavity dissipation
(decoherence) rates respectively \cite{strong}.
The symmetric collective internal states can reach the strong coupling
regime without requiring a high finesse cavity as
$g_0\propto \sqrt{N}$, with $N$ the
total number of atoms \cite{sf1,n3,n4}. When implemented with
protocols insensitive to individual atomic dissipation/decoherence
rate as in the Dark state based adiabatic transfer
protocols \cite{lukin0,lukin1,tom}, one can apparently gain as
upper hand over systems based on single atoms inside a cavity
\cite{Flei0,Flei1}. This is in fact, not quite surprising, earlier
cavity QED experiments have relied on the enhanced dipole
interaction of a collection of many atoms \cite{sf1,n3,n4}. In
free space, the phenomena of superfluorescence or super-radiance
\cite{sf1,sf} constitutes another example of collective state
dynamics. Recent experimental success clearly demonstrates the
power of such an atomic ensemble based system for entangling
macroscopic objects \cite{alex,polzik}. Several new ideas have raised
further expectation of exciting developments to come \cite{zol1,atac}.
 Nevertheless, all ensemble based systems suffer from
 the reduced size of computational Hilbert space. In this case
of symmetric collective internal states, the space used for
quantum information, is much less than the ${\cal V}^T=2^N$ as for
$N$ two level atoms \cite{julio}.
In view of the recent experimental success in
storage and recovery of light coherence in atomic gases
\cite{hau,ron}, a related question to address is the sensitivity to
errors when collective spin states are as quantum memories.

In this paper, we investigate the decoherence for
a superposition of symmetric internal states of an atomic gas due
to its inhomogeneous coupling with external control fields.
Our study is motivated by the simple observation that the
symmetric states of an atomic ensemble spans the
computation space only if atoms can be manipulated
cooperatively, namely, the coupling of both the external manipulating field
and the environment surrounding the atomic ensemble should be homogeneous
such that the collective motion of the atomic ensemble can be
described by the collective quasi-spin operators. In essence the effect of
different spatial positions for atoms $1,2,\cdots$, and $N$, is ignored or
absorbed into each single spin operators. In reality, an optically thick
atomic ensemble suffers from inhomogeneous coupling to both
classical and quantum light fields, i.e. the coupling strength is
position dependent. Such a situation arises naturally for trapped
ions due to its center of mass motion. In this case, it is well
known that the loss of quantum coherence for a superposition of
internal state occurs. In this study, we refer such decoherence
effect as inhomogeneous decoherence.
We focus on introducing our technique and study the simplest
example of superpositions of collective
atomic Dicke states in this paper.
The consideration of Dark-state, polariton approach based
proposals \cite{Flei1} will be given in the future, as significant
complications arise when the quantum cavity field is included.

Our model constitutes of an ensemble of two-level atoms
described by Hamiltonian
\begin{eqnarray}
H=\sum_{k=1}^N[\frac 12\omega_a^{(k)}\sigma_z^{(k)}+\frac 12%
(g_0^{(k)}\sigma_{+}^{(k)}+h.c.)],
\label{mh}
\end{eqnarray}
where the $\sigma$'s are the standard Pauli matrices ($\hbar=1$),
and the different local coupling $g_0^{(k)}$ (for the $k$-th atom)
may be due to a cavity mode profile as common in tightly focused
cavities or when atomic motional wave packet is insufficiently
localized \cite{howard}.
$\omega_a^{(k)}=\varepsilon_a^{(k)}-\omega_L$, is the difference
between atomic energy $\varepsilon_a^{(j)}$ and the external near
resonant laser frequency $\omega_L$. For convenience, we further
abstract Eq. (\ref{mh}) into the compact form
$H=\sum_{k=1}^N\vec{B}^{(k)}\cdot \vec{\sigma}^{(k)}$, with real
parameters $B_\mu ^{(k)}$.

The symmetric collective spin space
${\cal V}^S$ of dimension $2J+1\ll 2^N$ ($J=N/2$) is spanned by
the collective angular momentum states
$\{|J,M\rangle,M=-J,\cdots,J-1,J\}$ of $J_\mu
=\sum_{i=1}^N\sigma_\mu ^{(i)}/2$ satisfying ${\lbrack J_\mu,J_\nu
]}=i\epsilon_{\mu \nu \zeta}J_\zeta$ and
$J_{x}^2+J_{y}^2+J_{z}^2=\hat J^2=J(J+1)$. $\epsilon_{\mu \nu
\zeta}$ is the symmetric permutation tensor. The $|J,M\rangle$
space can be generated from the ladder operator $J_{\pm}=J_x\pm
iJ_y$ according to \cite{Dicke54},
\begin{eqnarray}
|J,M\rangle
=\sqrt{\frac{(J-M)!}{(J+M)!(2J)!}}\,J_{+}^{J+M}|J,-J\rangle,
\end{eqnarray}
except we note that an arbitrary unimodular phasor
can be self-consistently included with
$J_{\pm}=\sum_{k=1}^Ne^{\pm i\theta_k}\sigma_{\pm}^{(k)}/2$
and
$|J,-J\rangle =|\!\downarrow,\downarrow,\cdots,\downarrow\rangle$.

For any realistic system, an inhomogeneous
distribution of the parameter $\vec B^{(j)}$ makes it impossible
to constrain the system dynamics within the subspace ${\cal V}^S$.
To facility further discussion denote $H=H_0+H_1$ with
$H_0=\sum_{k=1}^N\vec{B}\cdot \vec{\sigma}^{(k)}$ and
$H_1=\sum_{k=1}^N\vec{b}^{(k)}\cdot \vec{\sigma}^{(k)}$,
where $\vec{B}^{(k)}=\vec{B}+\vec{b}^{(k)}$ with
$\vec{B}=\sum_k\vec{B}^{(k)}/N$.
$H_0$ constitutes the intended coupling between the symmetric
collective spin states, while $H_1$ represents a source of inhomogeneous
decoherence. It causes
decoherence as it provides a direct coupling from the subspace ${\cal V}^S$
to its complement ${\cal V}^O$ in ${\cal V}^T$.
A quantitative measure for the unwanted coupling
$H_1$ is in terms of the leakage parameter.
Suppose initially the system is prepared in a
superposition of collective spin states $|\phi(0)\rangle\in {\cal V}^S$.
The intended dynamics governed by
$U_0(t)=e^{-itH_0}=\prod_{k=1}^Ne^{-it\vec B\cdot\sigma^{(k)}}$
leads to the resultant state
$|\phi(t)\rangle_0=U_0(t)|\phi(0)\rangle$,
still within the same subspace. The actual final state
is $|\phi(t)\rangle=U(t)|\phi(0)\rangle$
with $U(t)=\prod_{k=1}^Ne^{-it\vec B^{(k)}\cdot\sigma^{(k)}}$
generally will span more than ${\cal V}^S$.
The leakage can therefore be defined as
\begin{eqnarray}
\xi&&=1-|\langle\phi_{0}(t)|\phi(t)\rangle|^{2}.
\end{eqnarray}
$\xi=0$ corresponds to no leakage, while
$\xi\to 1$ indicates a complete loss of the system
coherence and population.

Denote $|\phi (0)\rangle \in {\cal V}^S$ as a
normalized state expanded in terms of $|J,M\rangle$,
\begin{eqnarray}
|\phi (0)\rangle=\sum_{M\ll N, {\rm or} M\sim N} c_M|J,M\rangle,
\end{eqnarray}
the overlap
\begin{eqnarray}
|\langle\phi_{0}(t)|\phi(t)\rangle|^{2}&&=
|\langle\phi(0)|U_{0}^{\dagger}(t)U(t)|\phi(0)\rangle|^{2}\nonumber\\
&&=\sum_{M}\sum_{M'}c_{M'}^*c_M O_{M'M}(t) \le 1,
\end{eqnarray}
becomes the focus of our study with
\begin{eqnarray}
O_{M'M}(t)&&\equiv\langle
J,M'|U_{0}^{\dagger}(t)U(t)|J,M\rangle\nonumber\\
&&=\langle J,M'|\prod_{k=1}^NO^{(k)}|J,M\rangle,
\label{ff}
\end{eqnarray}
$O^{(k)}=R^{(k)}
+i\vec{I}^{(k)}\cdot \vec{\sigma}^{(k)}$, and
\begin{eqnarray}
R^{(k)}&&=\cos Bt\cos B^{(k)}t+ (\hat{n}\cdot \hat{n}^{(k)})\sin Bt\sin B^{(k)}t,\nonumber \\
\vec{I}^{(k)} &&=\hat{n}\sin Bt\cos B^{(k)}t+\hat{n}^{(k)}\cos Bt\sin
B^{(k)}t \nonumber\\
&&+(\hat{n}\times \hat{n}^{(k)})\sin B\sin B^{(k)}t.
\end{eqnarray}
We have defined $\hat{n}=\vec{B}/B$ and
 $\hat{n}^{(i)}=\vec{B}^{(i)}/B^{(i)}$.

The evaluation of Eq. (\ref{ff}) is difficult as state $|J,M\rangle $
involves a symmetric permutation of all atoms
so that the $\prod_k$ factor can not be pulled outside
the inner product. Furthermore, $\prod_{k=1}^NO^{(k)}$ expands
into $2^N$ separate terms, involving asymmetric products
of $\vec{\sigma}^{(k)}$ of upto powers of $N$.
A similar product structure was found to be responsible for
decoherence in quantum measurement models \cite{suncp},
where the decoherence factor
(the overlaps of the final states of
detector or a environment) suppresses the off-diagonal element of its
reduced density matrix. In mathematical terms,
for a factorized state $|f\rangle=\prod_{k=1}^N |f^{(k)}\rangle$,
the overlap integral $\langle f|\prod_{k=1}^NW_{M'M}^{(k)}|f\rangle$
becomes
$\prod_{k=1}^N\langle f^{(k)}|W_{M'M}^{(k)}|f^{(k)}\rangle$,
which approaches zero in the limit of macroscopic $N$
as each factor $\langle f^{(k)}|W_{M'M}^{(k)}|f^{(k)}\rangle$
has a norm less than unity. To make a similar argument for the
present problem, we need to find an expression such that the
collective state $|J,M\rangle $ becomes factorized. Since we
are interested in obtaining the asymptotically valid results
in the limit of large $N$,
a short time approximation (small $t$) can not be simply
adopted. Following early discussions on atomic coherent states
\cite{Gilmor,ueda}, we introduce
\begin{eqnarray}
|\theta\rangle=\prod_{k=1}^N\frac 1{\sqrt{2}}(1+{e^{i\theta}\over 2}
\sigma_{+}^{(k)})|\!\downarrow\rangle
=\frac 1{2^{N/2}}e^{J_{+}e^{i\theta}}|J,-J\rangle,
\end{eqnarray}
a phase coherent state, that can be expanded
according to the number of excitations
\begin{eqnarray}
|\theta \rangle &&={\frac 1{2^{N/2}}}
\left[1+e^{i\theta}J_{+}+\cdots+{e^{in\theta}\over n!}J_{+}^n+\cdots
\right]|J,-J\rangle \nonumber\\
&&=\sum_{M=-J}^J{e^{i(J+M)\theta} \over {\cal N}_{JM}}
|J,M\rangle,  \label{ac}
\end{eqnarray}
where ${\cal N}_{JM}=\sqrt{{(J+M)!(J-M)!2^N}/{(2J)!}}$.
The inverse transformation gives
\begin{eqnarray}
|J,M\rangle =\frac{{\cal N}_{JM}}{2\pi}
\int_0^{2\pi}e^{-i(J+M)\theta }\,|\theta \rangle d\theta,
\end{eqnarray}
which helps to
evaluate Eq. (\ref{ff}) as
$O_{M'M}={\cal N}_{JM}{\cal N}_{JM'}o_{M'M}$
with the reduced overlap
\begin{eqnarray}
o_{M'M}&&={1\over 4\pi^2}\int_0^{2\pi}d\theta \int_0^{2\pi}d\theta'\nonumber\\
&&e^{-i(J+M)\theta}e^{i(J+M')\theta'}\prod_{k=1}^NG^{(k)}(\theta,\theta')
\end{eqnarray}
in a simple factorized form and
$$
G^{(k)}={\frac 1 2}\,_k\langle \downarrow\!|
(1+e^{-i\theta'}\frac{\sigma_{-}^{(k)}}2)O^{(k)}
(1+e^{i\theta}\frac{\sigma_{+}^{(k)}}2)|\!\downarrow\rangle_k.
$$
$|G^{(k)}|\le 1$ as both
$(1+e^{i\theta}\frac{\sigma_{+}^{(k)}}2)|\downarrow\!\rangle_k/\sqrt{2}$
and $(1+e^{i\theta}\frac{\sigma_{+}^{(k)}}2)|\!\downarrow\rangle_k/\sqrt{2}$
are normalized. This points to a strong physical argument
against rapid decoherence of collective spin state qubits.
The question to answer is now clearly how does $O_{MM}$ approach
$0$ due to inhomogeneous coupling.
If the coupling coefficients
$g_0^{(k)}$ and $\omega_a^{(k)}$ were constants
(independent of atom label $k$),
$O_{MM'}\equiv\delta_{MM'}$.

\begin{figure}
\includegraphics[width=3.25in]{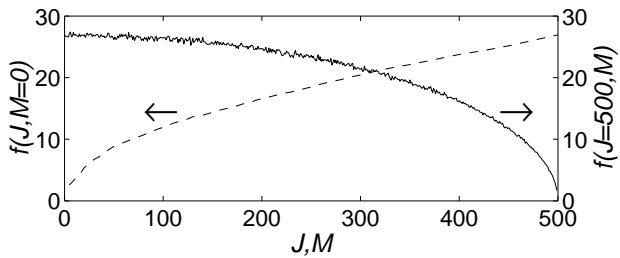}
\caption{$M$ and $J$ dependence of $f$ for $J=500$ and $B_z=1$.}
\label{fm}
\end{figure}

We investigate the above question for several model cases.
First, we look at inhomogeneous broadening
when $B_x^{(k)}=B_y^{(k)}=0$ and $B_z^{(k)}$
satisfies a normal distribution (with respect to $k$)
with mean
$\vec B=B_z\hat z=\langle B_z^{(k)}\rangle\hat z$,
and variance $\sigma_z^2$.
We find $O_{MM'}(t)\propto\delta_{MM'}$ with the
coefficient being a constant unity for
$|M|=J$ but decays with a time constant
$T_{1/2}\propto 1/(\sqrt{N}\sigma_z)$ for $|M|<J$ .
Define $T_{1/2}\equiv 1/({f\sigma_z})$, we find
$f$ is essentially independent of
$\sigma_z$ for $\sigma_z\in [10^{-7}, 10^{-1}]B_z$.
It contains an apparent dependence on $J^2-M^2$
as shown in Fig. \ref{fm} for a given $J$ and $B_z$.
The $J$ dependence (for $M=0$) is also shown in the same figure.
Based on our extensive numerical study, we find to a high
level of accuracy
\begin{eqnarray}
T_{1/2}(J,M,\sigma_z)=\frac{1}{\kappa\sigma_z\sqrt{J}\sqrt{1-M^2/J^2}},
\label{fit}
\end{eqnarray}
with $\kappa$ ($\approx 1.2$),
essentially independent of $B_z$ for $B_z\in[10^{-2},10^2]$.

\begin{figure}
\includegraphics[width=3.25in]{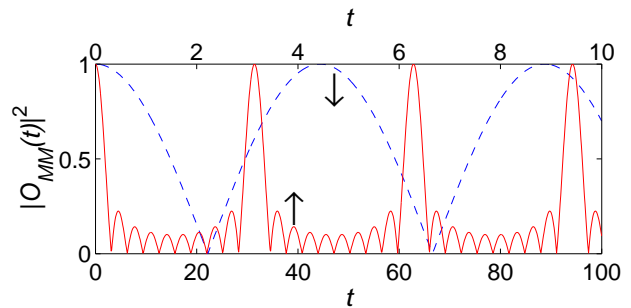}
\caption{Periodic behavior for $|O_{MM}(t)|^2$.
Solid line denotes $M=4$ and $N=10$ with respect to
the lower time axis, while dashed line denotes
$M=2$ and $N=2$; $B_z^{(k)}=k$ is taken for $N=10$
to assure the appearance of revival. For $N=2$
revival occurs for arbitrary random values of $B_z^{(k)}$.}
\label{period}
\end{figure}

This result is to be expected based on the collapse and revival of
a quantum wave packet \cite{revival}, since each individual atom
collapses with a time constant $\propto 1/\sigma_z$,
the collective states of an Guassian
ensemble should collapse with a time constant $\propto
1/(\sqrt{N}\,\sigma_z)$ as the net variance simply adds.
This is indeed what we find for $M=0$ or in general for
$|M|\ll J$. Equation (\ref{fit}) also indicates that
significantly reduced decoherence does occur in this case for
$|M|\sim J$, a regime where collective spin states are
mostly useful \cite{lukin0,lukin1,Flei0,Flei1,zol1,atac}.
In fact, for a single qubit quantum memory
involving the two state superposition of $M=-J$ and $-J+1$,
the decoherence rate is just that of a single atom \cite{Flei0}.
For small
values of $N$, when ratios of different coupling strength $B_z^{(k)}$
match ratios of integers, we indeed were able to
find the expected revival as shown in Fig. \ref{period}. This of
course will not happen for an ensemble with a macroscopic $N$.

\begin{figure}
\includegraphics[width=3.25in]{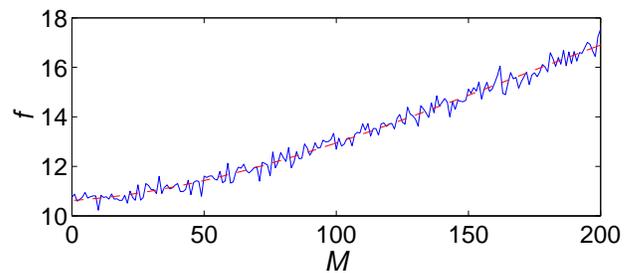}
\caption{The $M$ dependence of $f$ for $J=200$ and $B_r=10$.
The smooth curve is a fit given by
$-4.06483\times 10^{-7}|M|^3+2.03393\time 10^{-4}M^2
+0.00697|M|+10.62188$. } \label{xyfm}
\end{figure}

Next we consider the case of inhomogeneous Rabi coupling with
$B_{x/y}^{(k)}$ being Gaussian distributions with mean
$B_x=B_y=B_r$ and variance
$\sigma_x^2=\sigma_y^2=\sigma_r^2$, and $B_z^{(k)}=0$.
Similar to the previous case, we find the diagonal
term $O_{MM}(t)$ (including $M=\pm J$) decays with
a time constant $T_{1/2}=1/f\sigma_r$.
The $J$ dependence of $T_{1/2}$ is in fact
almost identical, i.e. $f_{M=0}=\kappa_1 J^{1/2}$,
with $\kappa_1\approx 0.76$ when $B_r=10$.
The $M$ dependence, on  the other hand is more
complicated as shown in Fig. \ref{xyfm}.
Obviously $f$ does not depend on $M$ linearly
as now $|O_{MM}(t)|$ seems to decay faster for
larger values of $|M|$.

\begin{figure}
\includegraphics[width=3.25in]{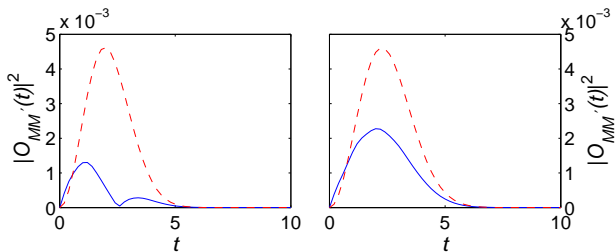}
\caption{Typical sampling of $O_{MM'}(t)$ for $J=100$,
$M=J-1$, and $M'=M-1$ (solid line) and $M-2$ (dashed line).
Different figures correspond to different random number sets.}
\label{offt}
\end{figure}

The off-diagonal element $O_{M\neq M'}(t)$
grows to significant nonzero values, as shown
by the typical sampling of $|O_{MM'}(t)|^2$
in Fig. \ref{offt} when $N$ is not too large.
Overall, we find
the dependence on the random number sampling
is strong only when $M-M'=\pm 1$,
so we focus on $M-M'=\pm 2$ here.
Define $T_{\rm max}$ as the time for $|O_{MM'}(t)|$
to reach its first maximum and ${\cal O}_{\rm max}$
the value of the maximum. We find that similar to
the diagonals, $T_{\rm max}=1/f\sigma_r$,
with $f$ a function of $J, M, M'$, and $B_r$,
although ${\cal O}_{\rm max}$ seems to be largely
independent of $\sigma_r$.
To study the $J$ dependence of $f$ and ${\cal O}_{\rm max}$,
we consider the limiting case of $|M|\sim J$
when collective states are usual proposed to work.
The result $f\propto J^{1/2}$ is once again as expected.
In this case, we also find quite accurately
${\cal O}_{\rm max}\propto J^{-1}$.

To summarize, we find within our model, the apparent
decoherence or dissipation rate for superpositions of
collective spin states scales as $\sqrt N$.
This evidence clearly demonstrates that asymptotically
there is no advantage of using collective spin states
for quantum information processing. The $\sqrt{N}$ enhanced
coherent dynamics is simply being compensated by the
$\sqrt{N}$ enhanced decoherence when inhomogeneous
coupling arises.

Finally, we note that our result also applies to the
case of entangled states between the collective spins of
two separate ensembles. For instance,
for two ensembles $A$ and $B$, a state
$\sum_{M_A,M_B}c_{M_A,M_B}|J_A,M_A\rangle_A|J_B,M_B\rangle_B$
can always be expressed as coherent superposition of
the total angular momentum basis $\vec J=\vec J_A+\vec J_B$,
i.e. into collective basis $|J_A,J_B;J,M=M_A+M_B\rangle$
with $J=J_A+J_B=(N_A+N_B)/2$.

We thank Mr. Yueheng Lan for some helpful discussion.
This work is supported by a grant from NSA, ARDA, and DARPA
under ARO Contract No. DAAD19-01-1-0667, and by a grant
from the NSF PHY-0113831.  Partial support
 from the NSF of China is also acknowledged.

\end{document}